\documentclass[sigconf,nonacm]{acmart}

\AtBeginDocument{%
  }

\usepackage{url}
\usepackage{soul}
\usepackage{xcolor}
\usepackage[dvipsnames]{xcolor}
\usepackage[T1]{fontenc}

\newboolean{showChanges}
\setboolean{showChanges}{False} %

\ifthenelse{\boolean{showChanges}}{%
 
\newcommand{\oldtext}[1]{\sout{#1}}
}{
\newcommand{\oldtext}[1]{}
 
}
\usepackage{graphicx}
\usepackage{enumitem}
\usepackage{subcaption}
\usepackage{setspace} 
\usepackage{booktabs}
\usepackage{siunitx}
\usepackage{makecell}
\usepackage{tabularx}

\usepackage[backgroundcolor=white, textcolor=magenta, linecolor=magenta, bordercolor=white, textsize=tiny]{todonotes}

\newcommand{\deletes}[1]{}

\begin{document}

\title{AI-Mediated Self: How HCI Defines and Relates to the Self}
\author{Jenny Xiyu Fu}
\email{xf89@cornell.edu}
\affiliation{%
  \institution{Cornell University}
  \city{Ithaca}
  \state{New York}
  \country{USA}
}

\author{Qian Yang}
\email{qy242@cornell.edu}
\affiliation{%
  \institution{Cornell University}
  \city{Ithaca}
  \state{New York}
  \country{USA}
}

\author{Malte Jung}
\email{mfj28@cornell.edu}
\affiliation{%
  \institution{Cornell University}
  \city{Ithaca}
  \state{New York}
  \country{USA}
  }

\renewcommand{\shortauthors}{Fu et al.}

\begin{CCSXML}
<ccs2012>
   <concept>
       <concept_id>10003120.10003121.10011748</concept_id>
       <concept_desc>Human-centered computing~HCI theory, concepts and mode</concept_desc>
       <concept_significance>500</concept_significance>
       </concept>
 </ccs2012>
\end{CCSXML}

\ccsdesc[500]{Human-centered computing~HCI theory, concepts and models}

\keywords{human-computer interaction, self, scoping review}

\begin{abstract}
How might AI alter how we understand and experience the self? This scoping review analyzes 102 papers to examine how the self is defined in the field of human-computer interaction (HCI), how AI--self relationships are conceptualized, and what risks emerge when AI becomes entangled with selfhood. Our synthesis makes three contributions. First, we define \textit{AI-mediated self} as a conceptual umbrella that connects dispersed work across education, workplace, health, and creative practices. Second, we consolidate six framings of the self with four domains of ethical risk---agency/autonomy, identity/authorship, relational capacity, and meaning-making---into a conceptual map that provides a reusable vocabulary across contexts. Third, we introduce the \textit{Inclusion of AI--Self} framework, which situates AI--self relationships along a spectrum of proximity. Together, these contributions position selfhood as a central design space in HCI.

\end{abstract}

\begin{teaserfigure}
  \centering
  \includegraphics[width=0.85\textwidth]{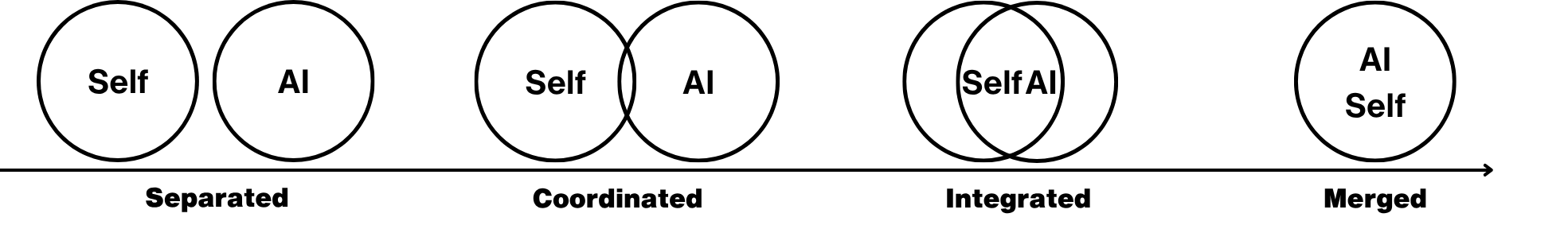}
  \caption{\textbf{The Proximity Framework for AI--Self Relations}, surfaced in our scoping review of 102 HCI papers from 2013 to 2025. Increasing overlap between the two circles marks increasing proximity between self and AI, from Separated to Merged.}  
  \Description{Four pairs of circles labeled Self and AI with increasing overlap, from Separated to Merged.}
  \label{fig:teaser}
\end{teaserfigure}

\maketitle
\section{Introduction}
The self stands as one of humanity's most complex concepts. To be human is to continuously construct narratives into coherent accounts of who we are, how we connect, and who we might become. If our sense of self is constructed from our memories of the past and plans for the future, what will happen if AI distorts them? 

Today, Generative AI represents the latest, perhaps the most invasive, technology in reshaping the self. Personalized AI systems now have the potential to generate content that allows users to integrate into their self-narratives, blurring the boundary between self-generated and system-generated expression. These interactions extend into increasingly personal parts of people's lives. For example, the emerging cases of AI psychosis highlight how LLM chatbots are changing people's self-perception and awareness~\citep{carnevale_studies_nodate}. Beyond the individual, people are developing intimate bonds with AI companions like Replika or interacting with AI ``clones'' of deceased loved ones~\citep{adam_supportive_2025,worth_ready_2025}. As a consequence, the self increasingly emerges through interaction with technical systems, part autobiographical, part computational, continuously co-constructed with algorithmic artifacts. This subtle but pervasive shift has raised attention in both public discourse (e.g., discussions on AI existential risks)~\citep{dean_does_2025,noauthor_generative_nodate} and academic research (e.g., the call for papers on Agency and Identities for ACM CUI 2026~\citep{noauthor_full_nodate}). With the explosive growth in this area, it is clear that AI is reshaping people's sense of self in unprecedented ways. As HCI researchers, we urgently need to establish better understanding of AI-self interactions. In this paper, we raise three central research question:

\begin{itemize}[leftmargin=*]
    \item How has HCI research defined self in HCI?
    \item How has HCI research conceptualized the relationships between AI and self?
    \item How has HCI research identified risks in AI-self interactions?
\end{itemize}

We presented a scoping review of 102 publications from ACM SIGCHI venues, the International Journal of Human-Computer Interaction, and the Human-Computer Interaction journal from 2013 to 2025, and our analysis of these publications identified six distinct framings of the self, four types of AI-self relations, and four categories of risks that appear across the literature. We made three contributions:
\begin{itemize}[leftmargin=*]
    \item A glossary of self-related definitions in HCI;
    \item A framework mapping AI-self relations, including how AI systems shape, interact with, or participate in self-related processes;
    \item A set of design recommendations for system designers, ethics researchers, and behavioral scientists based on observed risks and harms.
\end{itemize}

We concluded the paper with opportunities for future research, where we see understanding the self in relation to AI systems as essential for building resilience against GenAI's tendency of distorting the self. To shape a more desirable future, we need to understand how researchers currently study and design these systems, map the existing and emerging design approaches, and provide researchers and designers with frameworks to navigate this space thoughtfully.

\section{Background}
The study of self is fundamental to HCI because it shapes how users interact with, interpret, and integrate technology into their lives. From avatar customization to social media filter selection, HCI design decisions directly impact users' sense of agency. As interactive systems become more personally adaptive and socially embedded, understanding self-concept becomes crucial for designing ethical experiences that empower users’ psychological resilience and identity coherence. We clarify our motivation in Section 2.1, offer our definition of AI in Section 2.2, and provide a review of the HCI tradition on self research in Section 2.3.

\subsection{Why Self? Why Now?}
Our scoping review examines the nature of research and identifies gaps in the existing HCI literature on AI and self, which at its broadest, refers to an individual's internal concept of who they are \cite{epstein_self-concept_1973}. Self plays a critical role in motivation, emotion regulation, and social interaction. It provides continuity across time, helping individuals make sense of their past, plan future actions, and build a coherent identity. The self-concept supports essential psychological functions, including life story continuity, value-consistent behavior, and the foundation for social relationships. Since the study of self is inherently multifaceted, to develop a comprehensive search strategy that would not be constrained by any single disciplinary lens, we surveyed these related fields to identify key constructs and these theoretical references directly informed our search term selection (Section 3.2).

AI and self is an emerging area in HCI. Previous work in personal informatics, robotics, and XR shows how technologies mediate people's self-understanding, and current GenAI applications has shifted interactive systems from tools to co-authors, moving questions of ownership and agency from theoretical considerations to immediate design challenges. Public discussion of terms such as ``AI psychosis~\citep{hudon_delusional_2025}'' and ``rent a human for your AI~\citep{noauthor_man_nodate}'' indicates growing attention to these systems, and people are already discussing the unintended harm and consequences. Research addressing these phenomena has emerged across multiple domains, including education, healthcare, and creative practice, but findings remain fragmented across different subfields and theoretical traditions. This review synthesizes 102 studies to assemble shared terms and map recurring understanding of the self. Our position is that how HCI characterizes AI-self relationships will influence both research and the systems that are built. 

\subsection{What is AI?}
Defining AI is difficult as its boundaries shift from yesterday's breakthroughs to today's infrastructure. While ``AI'' was once a relatively niche technical term~\citep{khosrow-pour_dba_history_2009}, it is increasingly being adopted into different forms of user-facing interactive systems~\citep{yang_re-examining_2020}. Advances in large language models have catalyzed widespread public interest in AI, drawing researchers beyond the traditional AI research community to include practitioners and researchers from philosophy, design, and social sciences. HCI, as an inherently interdisciplinary field, is particularly positioned to engage with this complexity. Previous work in human-AI interaction has noted this challenge: the technical boundary of AI is disputed even within AI research communities and continuously evolving~\citep{raji_ai_2021}. 

\subsubsection{Scope of AI in This Review}
\label{sec:ai-scope}
For this review, rather than adopting a universal definition, we scope ``AI'' functionally. AI refers to interactive and user-facing systems that participate in self-related activities. We consider both implemented systems and explicitly envisioned systems discussed in conceptual or design-oriented research. Their relevant functions include generating content, conducting dialogue, making predictions or classifications, and recommending or personalizing information. This scope includes generative AI and large language models, conversational agents, recommender systems, and intelligent assistants supporting cognition and creativity. We exclude backend automation and AI used only for infrastructure. Thus, our conclusions concern the literature identified within this operational scope and the selected publication sources. The search strategy and its coverage limitations are reported in the Methods.

\subsection{HCI tradition on Self Research}
HCI's study on selfhood has grown through waves of technologies, with each wave importing different theories, units of analysis, and design goals. In the sections below, we trace each tradition to surface what HCI collectively knows about technology and the self.

The beginning of HCI research on the self emerged from studies of early online communities and virtual environments. Goffman's theory~\citep{goffman_presentation_2023} of self-presentation became a foundational framework, with researchers examining how interface design shapes the ``stages'' on which users perform different aspects of selfhood. Turkle's work~\citep{turkle_cyberspace_1999} on online multiplicity revealed how users experimented with different aspects of identity across digital contexts, challenging assumptions about unified, stable selfhood. Research on multi-user dungeons and early social platforms demonstrated that digital environments afforded new forms of identity play and self-exploration that were previously unavailable in offline contexts. These early-stage works ground the foundation of design lessons on how different modality enables selective self-presentation, and designers should provide users with flexible tools to experiment with their identity.

The emergence of personal informatics and self-tracking technologies marked the progression in HCI's engagement with selfhood. Research in this domain has examined how self-monitoring technologies shape self-awareness, behavior change, and personal meaning-making~\citep{epstein_mapping_2020,elsden_quantified_2016}. The quantified self movement introduced technologies that promised to reveal users’ ``true self'' through data collection and analysis. However, studies of fitness trackers, mood trackers, and life-logging systems have also revealed more complex dynamics~\citep{carrington_but_2015,rawassizadeh_ubiqlog_2013,adams_towards_2014}. Rather than straightforward self-improvement, these technologies often generated anxiety and self-criticism. To support better self-awareness in context, personal informatics researchers have used design principles such as personalized visualizations for self-tracking systems, social feedback to enable community engagement, and embedded tracking practices to foster collective support.

At the same time, another strand of HCI research has explicitly positioned self-development as a design goal. This approach drew on positive psychology and self-determination theory to create frameworks for evaluating whether technologies support or undermine psychological well-being and authentic self-expression. In Zimmerman's work, ``design for the self'' introduced a design focus on helping people become the person they desire to be through product interactions.~\citet{zimmerman_designing_2009}'s approach is grounded in product attachment theory, which describes how people develop emotional connections to possessions through meaning-making and identity construction. Unlike static objects that people select to represent who they are, interactive technologies can actively participate in identity formation through their behavioral responses and adaptive capabilities. Drawing on findings to support a better self, HCI researchers have designed artifacts to support better self-paced learning experiences by aligning course design to users’ values, and customization in mixed-reality.

In the critical and humanist HCI space, researchers drew on~\citet{foucault_history_1977}'s ``care of the self'' to examine how digital technologies mediate self-reflections and therefore shape who they are. Critical HCI theories challenged previous HCI assumptions about stable and coherent selfhood by demonstrating that technology and design are not neutral, and they create power and exacerbate existing societal issues, thus producing impacts that transform users' identities For example, ~\citet{bardzell_lonely_2014}'s research on Second Life revealed how avatar design functions as a ``technology of the self,'' where users engage in askēsis (labored self-consideration) through practices such as gender exploration and intimate role-play that blur the boundaries between the user and their digital representation.~\citet{ahman_conceptualizing_2017}'s critical analysis of ``Conceptualizing the Self'' draws on theories by Mark C. Taylor, using poststructuralism to reframe technology from a user-tool dynamic to what Taylor calls ``existential co-constitution,'' where interactive technology becomes an ``existential arena'' and interaction itself an ``existential process'' that constitutes selfhood. To surface power dynamics and translate insights into practices, researchers leverage design methods, including participatory design, fieldwork, and longitudinal observations, to closely examine how users navigate their agency in sociotechnical ecosystems.

While this diversity reflects genuine differences in technological affordances and design contexts, it creates a practical problem as AI systems become increasingly general-purpose. To examine what HCI collectively understands about technology and selfhood, we are motivated to conduct the literature review presented in the current paper.

\section{Method}
We conducted a scoping review to map how HCI research defines the self, conceptualizes relationships between AI systems and the self, and characterizes risks arising from those relationships. To ensure our search terms were theoretically comprehensive and captured the full range of self-related constructs across different research traditions, we identified working search terms from social science and humanities literature. We report the review following the PRISMA-ScR (Preferred Reporting Items for Systematic reviews and Meta-Analyses extension for Scoping Reviews) guidelines~\citep{triccoPRISMAExtensionScoping2018}. In the following section, we describe the information sources, the boundary of the review, eligibility criteria, screening process, data charting, and analysis.

\subsection{Information Sources}
We purposively selected publication sources that cover general HCI research and subfields in which user-facing AI and self-related questions commonly appear, following the source-selection approach of prior HCI reviews~\citep{tang_llm_2025,pang_understanding_2025,epstein_mapping_2020}. This include 12 conference and proceedings sources and 3 leading journals: CHI, CSCW, DIS, IUI, UIST, HRI, CUI, HAI, UbiComp, UMAP, Creativity \& Cognition (C\&C), and Interaction Design and Children (IDC), ACM Transactions on Computer-Human Interaction (TOCHI), the \textit{International Journal of Human-Computer Interaction}, and \textit{Human-Computer Interaction}. We considered additional engines (e.g., Google Scholar, Scopus), but opted not to include them due to scale (e.g., ``AI and Self'' returns 6,760,000 articles on Google Scholar). This source selection creates a bounded map of HCI research and our findings should therefore be read as patterns in the selected HCI sources, not as estimates of all research on AI and the self.

\subsection{Eligibility Criteria}
We assessed eligibility along three separate dimensions: publication scope, AI involvement, and engagement with the self. Separating these decisions was important because a paper could concern a user-facing AI system without examining the self, or could mention a self-related term while using ``AI'' only for a backend or technical component. Table~\ref{tab:eligibility-criteria} states how we applied each criterion. Papers were retained only when they met all three.

\begin{table*}[t]
\caption{Eligibility criteria and examples that mark their boundaries.}
\label{tab:eligibility-criteria}
\centering
\small
\setlength{\tabcolsep}{4pt}
\renewcommand{\arraystretch}{1.08}
\begin{tabularx}{\textwidth}{@{}p{0.13\textwidth}X X X@{}}
\toprule
\textbf{Criterion} & \textbf{Included} & \textbf{Excluded} & \textbf{Boundary case} \\
\midrule
\textbf{Publication scope} &
Peer-reviewed archival research articles published in the selected HCI sources by July 10, 2025. Empirical, conceptual, and design-oriented papers were eligible. &
Non-archival items, papers outside the selected sources, and records published after the search date. &
A technical or conceptual paper was eligible only if it also met the AI-involvement and self-engagement criteria. \\
\addlinespace
\textbf{AI involvement} &
A user-facing system retrieved through our AI-related terms that performed a conversational, generative, predictive, classificatory, or personalized function in direct interaction with people. &
AI used only for backend automation, infrastructure, system optimization, or algorithmic performance evaluation without direct relevance to people's interaction or experience. &
A rule-based chatbot could be included when it was treated as an intelligent interactive agent and participated directly in a self-related practice. \\
\addlinespace
\textbf{Engagement with the self} &
At least one self-related construct informed the paper's research question, theoretical framing, design goal, measure, analysis, or central finding. &
``Self'' used only technically, such as self-driving or self-supervised learning, or identity, agency, and related constructs mentioned only incidentally. &
A paper using self-efficacy as a central outcome was included; a paper mentioning that a system might ``increase agency'' only as a general benefit, without developing or examining the construct, was excluded. \\
\bottomrule
\end{tabularx}
\end{table*}

We applied the operational scope defined in Section~\ref{sec:ai-scope}. This scope allows comparison across changes in the technologies and terms used from 2013 through 2025. Inclusion in the review does not imply that rule-based chatbots, recommender systems, predictive models, and large language models have equivalent capabilities. It indicates that the source paper positioned the system within the review's user-facing AI boundary and connected it substantively to the self.

\subsection{Search Strategy}
\label{sec:search-strategy}
We developed the search query through three pilot rounds before conducting the final search. In the first round, we found a broad query combining ``AI'' and ``self'' returned 36,252 records in the ACM Digital Library and produced many technical false positives, including self-driving and self-supervised learning. In the second round, we specified the self-related terms using constructs drawn from cognitive science, psychology, and sociology~\citep{gallagher_oxford_2011,cooley_looking-glass_1902,bem_self-perception_1972,mcadams_psychological_2013,james_consciousness_1890,rosenberg_rosenberg_1965,markus_cultural_1991,goffman_presentation_2023,jones_toward_1982}. We then validated this strategy through citation searches on key papers and confirmed adequate retrieval of relevant literature. This substantially reduced false positives while capturing the multidimensional nature of self as defined. We then combined the self-related terms using \textit{OR} so that a paper could be retrieved through any one of the selected construct. Furthermore, we tested broader variations (e.g., adding ``personhood,'' ``subjectivity,'' ``autonomy'') but found they either yielded redundant results already captured by our existing terms or introduced false positives falling outside our working definitions of self-constructs. In the third round, we expanded the AI-related terms to account for changes in HCI terminology over time. We found that ``AI'' alone missed papers using domain-specific terminology: HCI researchers often use ``chatbot'' or ``conversational agent,'' NLP papers use ``language model,'' and post-2022 work increasingly employs ``GPT,'' ``LLM,'' or ``generative AI.'' 

During query development, we used previously known relevant papers as sensitivity checks. After each revision, we checked whether the query retrieved these papers and manually examined samples of newly retrieved records for false positives. The final conceptual query connected the following two term sets with \texttt{AND}:

\begin{quote}
\begin{sloppypar}
\noindent{\ttfamily\textbf{AI Terms:} ["artificial intelligence"] OR ["ai"] 
OR ["language model"] OR ["generative ai"] OR ["gpt"] OR ["chatgpt"] OR 
["large language models"] OR ["llm"] OR ["natural language processing"] 
OR ["nlp"] OR ["chatbot"] OR ["conversational agent"]]}

\medskip
\noindent{\ttfamily\textbf{Self terms:} [["self"] OR ["identity"] OR 
["selfhood"] OR ["self-perception"] OR ["self-concept"] OR ["self-efficacy"] 
OR ["sense of self"] OR ["agency"] OR ["personal identity"] OR 
["self-awareness"] OR ["self-understanding"]]}
\end{sloppypar}
\end{quote}

We conducted the final search on July 10, 2025 and included work published by that date. Although computer-vision systems were eligible under our operational scope, we did not include dedicated task terms such as ``image recognition'' or ``facial recognition.'' In the corpus, relevant examples like papers on AI-generated self-portraits were retrieved through terms such as ``generative AI.'' 

We acknowledge heterogeneity within this scope. Our corpus includes simple rule-based chatbots deployed in controlled lab studies, sophisticated LLMs integrated into commercial applications, generative AI installations in museums, and everything in between. We deliberately maintain this inclusive approach because HCI research on self and identity spans this full spectrum of AI systems. We validated each addition by checking whether it retrieved relevant papers we had previously identified through manual searches. Where relevant, our analysis accounts for meaningful differences in system capabilities, deployment contexts, and interaction paradigms.

\subsection{Overview of the Final Corpus}
\begin{figure}[t]
    \centering
    \includegraphics[width=\columnwidth]{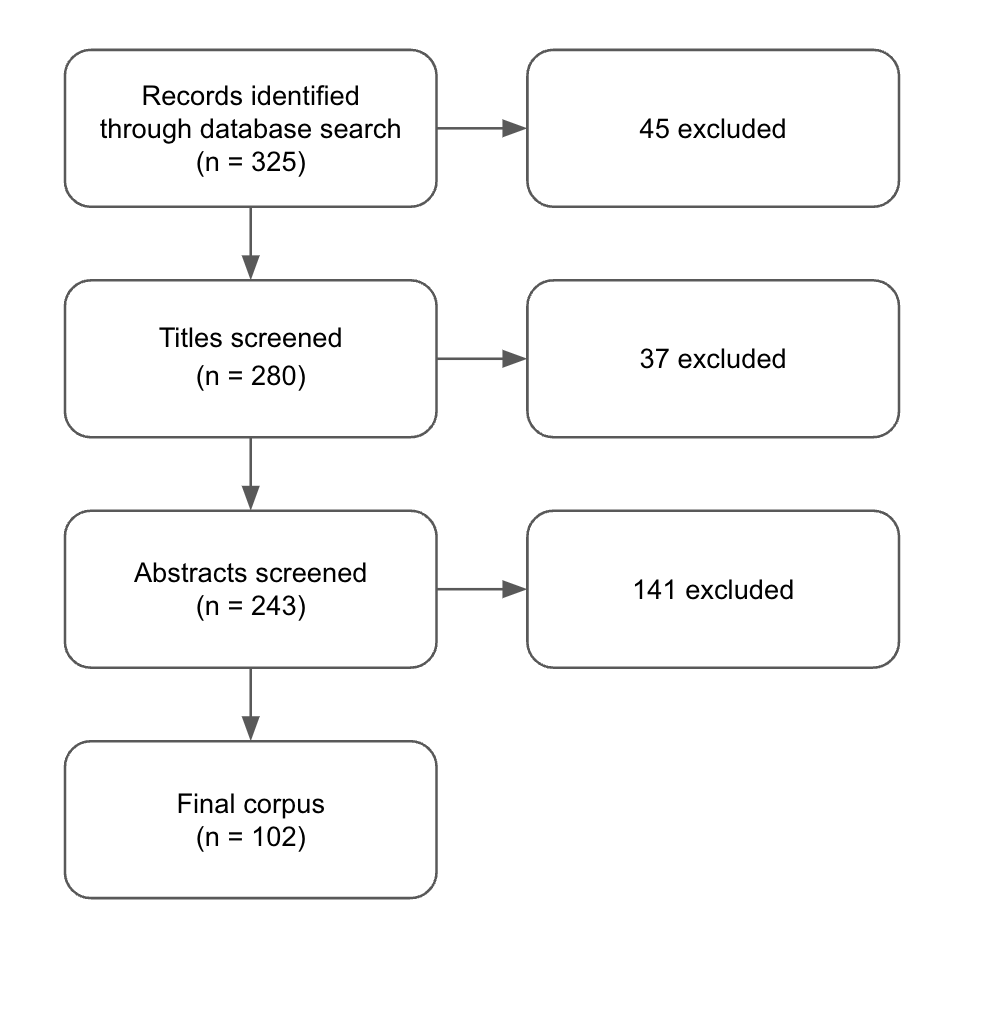}
    \caption{PRISMA Flow Diagram for Screening.}
    \Description{PRISMA-ScR flow diagram: 325 records identified, 45 removed before screening, 280 titles screened, 243 abstracts assessed, and 102 papers included.}
    \label{fig:PRISMA}
\end{figure}

The searches identified 325 records before screening. Following the PRISMA-ScR guideline\cite{tricco_prisma_2018}, we removed 45 records before title screening because they were duplicates, incomplete records, or papers for which the full publication could not be retrieved. We screened the remaining 280 titles and excluded 37 records that clearly used ``self'' in a technical sense, fell outside the selected publication scope, or did not concern user-facing AI. We then assessed 243 abstracts against the full eligibility criteria. The first and final authors reviewed these abstracts independently, discussed disagreements, and consulted the full text when the abstract did not provide enough information. This stage excluded 141 records. The first author then conducted a full-text confirmation of the remaining 102 papers, all of which were retained for analysis. The final corpus therefore consisted of 102 papers and spanned publications from 2013 through July 2025. Figure~\ref{fig:PRISMA} presents the records at each stage. 

\subsection{Data Extraction and Coding}
The paper was the unit of analysis. We aimed to analyze 102 papers consistently, despite the fact that many of these papers discussed the concept of `self' and `AI' abstractly and vaguely. We addressed that by (1) immersive reading and data extraction by the first author, (2) collaborative codebook development through iterative team discussions, and (3) thematic synthesis to construct analytic narratives. The first author read the full text of all 102 papers across three analytic passes. For each paper, the first author recorded descriptive information including research context, population, research question, method, and AI system, and wrote structured summaries addressing each review question. These paper-level summaries formed the initial coding material.

Using the first version of the codebook, the first author coded a sample of 20 papers and brought the coded cases and supporting excerpts to the full research team. The three authors then met weekly for 11 weeks to compare interpretations, examine uncertain and boundary cases, revise code definitions, and distinguish overlapping concepts. The codebook included four core dimensions: Self Aspect, Self Definition, AI System, and AI Role. We also recorded descriptive fields: Context, Population, Study RQ, and Method. The first author then applied the revised codebook across the full corpus.

Each paper could receive multiple codes within a dimension. In the findings, $n$ denotes the number of papers assigned to each category. A paper is counted once within each category but may contribute to several categories. For self-related concepts (RQ1), we extracted explicit definitions where available. When papers did not explicitly define the self, we interpreted how they conceptualized it through their theoretical framing, related work, and measures used to operationalize self-related constructs. For AI roles and AI-self relationships (RQ2), we drew on authors' descriptions of their systems and interactions, together with their reported analyses, to interpret how AI was positioned in relation to the self. For consequences and concerns (RQ3), we extracted reported outcomes and authors' accounts of potential risks and harms, which appeared primarily in discussion sections. Throughout coding, we extracted supporting quotations to ground our interpretations in the source papers. Fields were left blank when the text provided insufficient evidence to support either direct extraction or an evidence-grounded interpretation.

\subsection{Thematic Synthesis and Framework Development}
To make sense of this interdisciplinary and methodologically diverse body of work, we adopted \textbf{thematic synthesis} as our analytic approach. Thematic analysis allows systematic integration of findings across heterogeneous study types, ranging from qualitative empirical research, conceptual papers, and design-oriented work, which characterized the characteristics of our corpus~\citep{braunThematicAnalysis2012}. We are able to surface not only convergent themes but also tensions, contradictions, and blind spots across the literature. Moreover, since many papers discussed ``self'' implicitly, thematic analysis allow us to explicitly accommodates interpretive analysis of abstract or vaguely defined constructs. We did not calculate inter-rater reliability for this interpretive synthesis. This was also a practical decision, given the scope of the review, which included 102 full papers, each requiring multiple readings and in-depth theoretical interpretation. To support the trustworthiness of the analysis, the three authors examined supporting excerpts, discussed competing interpretations, reviewed ambiguous and boundary cases, and refined code definitions and higher-level themes through the 11 weeks of meetings~\citep{mcdonaldReliabilityInterraterReliability2019}.

\section{Findings}

\begin{figure}[t]
    \centering
    \includegraphics[width=\columnwidth]{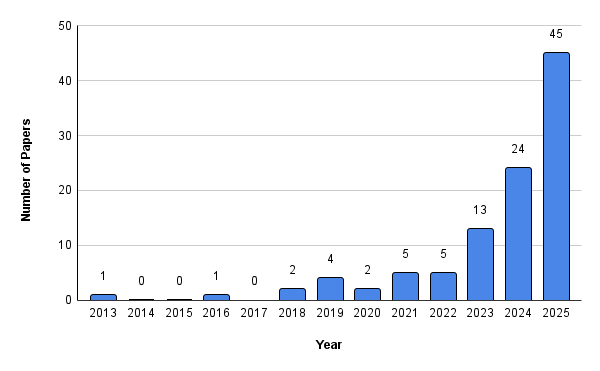}
    \caption{The number of self-related papers between 2013 and 2025.}
    \Description{Chart of the number of papers in the corpus per year from 2013 to 2025, rising sharply in 2024 and 2025.}
    \label{fig:time_chart}
\end{figure}

Our corpus comprises 102 peer-reviewed publications from HCI and related fields spanning 2013-2025, capturing the rapid emergence of interests in AI and selfhood. As Figure~\ref{fig:time_chart} illustrates, the temporal distribution reveals this as a distinctly contemporary phenomenon: nearly 70\% of papers were published in 2024-2025, with the earliest papers identified through the search dating to 2013. Table~\ref{tab:clusters} summarizes the major clusters we identified across the corpus. We follow our research questions to organize our findings.

\begin{table*}[h]
\centering
\setlength{\tabcolsep}{8pt}
\begin{tabular}{p{4.1cm} p{10.8cm}}
\toprule
\textbf{Code} & \textbf{Definition} \\
\midrule
\multicolumn{2}{l}{\textbf{RQ1 — Framings of the Self in HCI}}\\
Self as a dynamic process of Transformation (n=46) & The self as something shaped through reflection and ethical self-work, technologies scaffold ongoing change and growth.\\
Self as a Narrative Identity (n=30) & The self as an evolving life story, tools intersect with how people author past, present, and imagined futures. \\
Self as a Social Construct (n=49) & The self as situated in culture and relationships, systems mediate norms, roles, and interpersonal context. \\
Self as a set of Psychological Needs and Capabilities (n=59) & The self as motivation and skill, designs touch autonomy, competence, relatedness, self-efficacy, and self-awareness. \\
Self as a Performance and Act of Disclosure (n=48) & The self as performance to others, interfaces shape impression management and what people reveal. \\
Self as A Set of Values and Act of Meaning-Making (n=44) & The self as commitments and significance, interactions support purpose, coherence, resonance, and sense-making. \\
\midrule
\multicolumn{2}{l}{\textbf{RQ2 — Conceptualizations of AI–Self Relationship}}\\
AI as Bounded Cognitive Tool (n=69) & AI extends reasoning and workflow while remaining outside identity and under explicit human control. \\
AI as Collaborative Partner (n=88) & AI participates in co-creation and development, and people negotiate authorship, control, and interpretation. \\
AI as Intimate Mirror (n=54) & AI folds into inner dialogue, reflecting feelings, memories, and perspectives to prompt self-understanding. \\
AI as Representative Extension (n=24) & AI speaks or acts for the person in social settings, raising questions of authenticity and ownership. \\
\midrule
\multicolumn{2}{l}{\textbf{RQ3 — Ethical Risks in AI–Self Interaction}}\\
Agency and Autonomy (n=72) & Risks to real and felt control, from automation, design constraints, and over-reliance. \\
Identity, Authorship, and Self-Perception (n=62) & Substitution or distortion of voice and self-expression, bias, and stereotyping shape how people see themselves. \\
Relational Capacity and Emotional Regulation (n=57) & Dependency on agents, masking, displacement of human bonds, reduced psychological safety. \\
Confidence, Reflection, and Meaning-Making (n=48) & Mis-calibrated confidence, rumination and regret, disrupted continuity of personal meaning. \\
\bottomrule
\end{tabular}
\vspace{0.1cm}
\caption{Major clusters for RQ1, RQ2, and RQ3. A paper can receive multiple codes.}
\label{tab:clusters}
\end{table*}

\subsection{Six Framings That Define the Self in HCI}
\begin{figure*}[t]
  \centering
  \includegraphics[width=\textwidth]{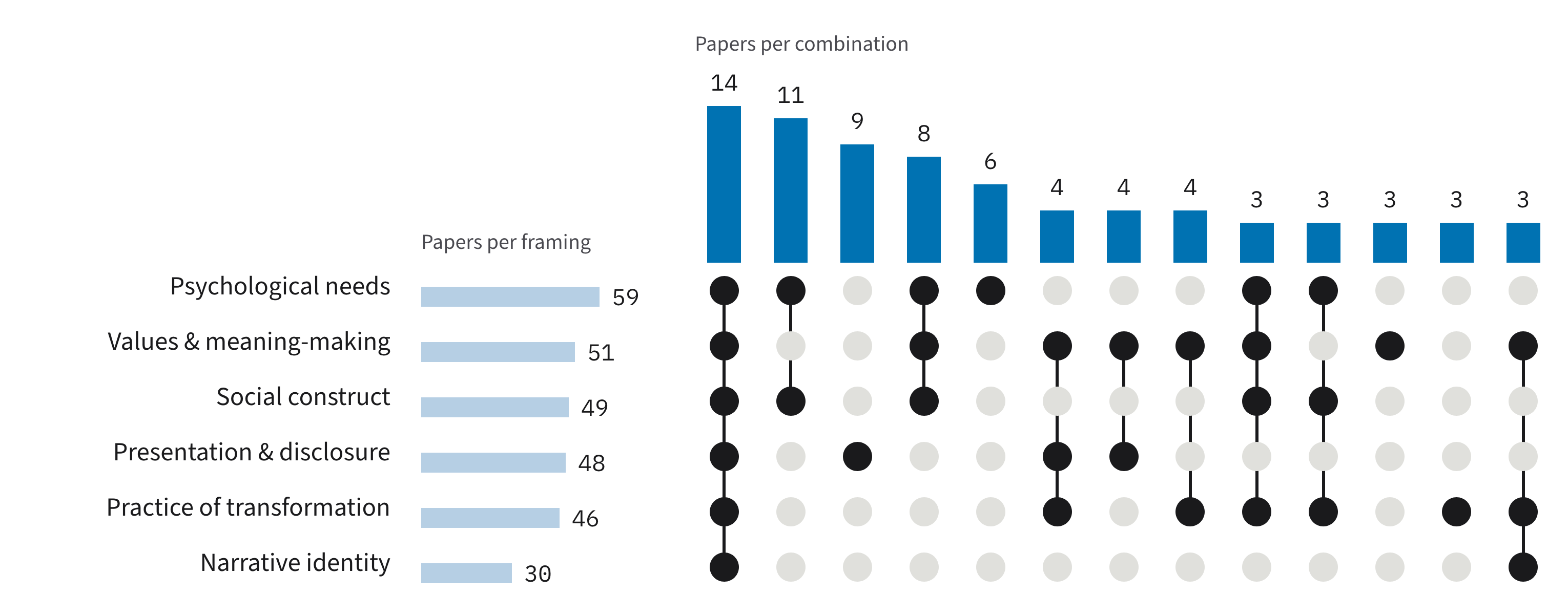}
  \caption{Frequency and co-occurrence of the six self framings. Horizontal bars show the total number of papers coded with each framing. Vertical bars show the number of papers with each exact combination indicated by the black dots below; gray dots indicate framings absent from that combination.}
  \Description{UpSet plot of six self framings and their co-occurrence across papers.}
  \label{fig:framing-upset}
\end{figure*}

Our synthesis identified six overlapping framings of the self. We connect them through a shared question: by defining the self differently, \textit{what would it mean for AI to change the self?} The framings direct attention to different aspects of this question: people's perceived abilities, their social relationships, their expression of themselves, their practices of meaning-making, their values, and their life stories. In the following section, we present the six framings in descending order of frequency within the corpus, describe its conceptual focus, connect the pattern to relevant theoretical traditions, and examine supporting examples. Figure~\ref{fig:framing-upset} shows the frequency and co-occurrence of these framings across the corpus. We also summarizes the methods used in papers associated with each framing in Table~\ref{tab:measurement}.

\subsubsection{The Self as Psychological Needs (n=59)}
When a user completes a difficult task with a AI system, what do they learn about their own capabilities? The most common framing conceptualized the self through psychological needs and motivation to act. To study the self as psychological needs, researchers use experimental behavioral science method like surveys and performance metrics. Papers also use psychological theory to guide system design. For example, to enhance self-efficacy, \citet{jeong_conversation_2025} designed a conversational progress guide with visual feedback like progress bars and subtask markers. Similarly, by applying psychological theories to support sense of belonging, projects like \citet{elwahsh_linguistic_2025} showed how custom multilingual chatbots can support emotion regulation and mental health for individuals when they are navigating multiple contexts.

Feeling more capable, however, does not necessarily mean assessing one's capabilities more accurately. In a contrasting example, \citet{he_knowing_2023} found that participants who overestimated their performance on reasoning tasks tended to under-rely on AI advice and lead to poorer human-AI performance. Supporting confidence is different from supporting accurate self-assessment, and this framing makes the instability raised in our introduction concrete: AI-related interactions can become part of how people assess what they know and what they can do. 

We relate these concerns to Self-Determination Theory's account of autonomy, competence, and relatedness~\citep{deci_self-determination_2012}. Albert Bandura's self-efficacy theory complements SDT by emphasizing belief in one's ability to achieve goals as a key motivational driver~\citep{bandura_self-efficacy_1982}. Related constructs include self-awareness, becoming the object of one's own attention, alongside self-disclosure and self-talk as regulatory and developmental processes~\citep{latinjak_self-talk_2023,bem_self-perception_1972,ballou_self-determination_2022}.

\begin{table*}[h]
\centering
\small
\setlength{\tabcolsep}{5pt}
\begin{tabular}{p{2cm} p{2.8cm} p{4.5cm} p{3.5cm}}
\toprule
\textbf{Framing} & \textbf{Methods} & \textbf{Scales} & \textbf{Data Types} \\
\midrule

Transformation
& Mixed-methods, diary studies, field deployments
& Authenticity Scale;
  PHQ-9/OASIS;
  Negative Automatic Thoughts (ATQ-N)
& Journey maps, diary entries, think-alouds \\

\addlinespace

Narrative
& Qualitative content analysis, mixed-methods with writing prompts
& LIWC;
  Self-Disclosure Depth (3-level coding);
  Pennebaker's Expressive Writing Questionnaire
& User-generated narratives, expressive writing, story completions \\

\addlinespace

Construct
& Participatory co-design, cross-cultural surveys
& Social Value Orientation (SVO);
  Inclusion/Control/Ownership scales;
  Mayer's Trust Model
& Workshop artifacts, cross-cultural comparisons, co-design outputs \\

\addlinespace

Capabilities
& Experimental, behavioral quantitative
& SDT scales (autonomy, competence, relatedness);
  Creative Self-Efficacy;
  NASA Task Load Index (NASA-TLX)
& Survey responses, task performance metrics, behavioral logs \\

\addlinespace

Presentation
& Online experiments, interaction logging, think-aloud
& Distress Disclosure Index (DDI);
  Ownership/Autonomy/Control;
  Perceived Trustworthiness
& Interaction logs, coded disclosure transcripts, text analysis \\

\addlinespace

Meaning
& Phenomenological, first-person, design explorations
& PANAS;
  Mekler \& Hornb{\ae}k meaning dimensions;
  Self-Congruence Scale
& Reflective writing, somatic responses, design artifacts \\

\bottomrule
\end{tabular}
\caption{Measurement by self-framing.}
\label{tab:measurement}
\end{table*}

\subsubsection{The Self as Values and Meaning-Making (n=51)}
How does AI change what is meaning for me? This framing conceptualizes the self as a locus of values and meaning-making, where technologies intersect with personal commitments and the search for significance. Methodologically, this cluster draws phenomenological inquiry and design explorations that foreground first-person experience and value-laden encounters with AI, alongside mixed-methods studies that assess constructs like self-congruence and emotional resonance. For example, \citet{wan_metamorpheus_2024} introduces Metamorpheus, an affective creative AI facilitating meaning-making through emotional resonance in creative expression. Additionally, the design space of ``introspective AI'' examines how AI can reveal tacit qualities and support self-knowing~\citep{lustig_explainability_2022,nunez-pacheco_searching_2025}. Projects like ``ML Tarot''~\cite{lustig_explainability_2022} demonstrate how randomness and uncertainties of AI can help designers gain deeper self-understanding.

We connect this attention to personally meaningful experience with studies of introspective AI and self-knowing. Theories of self-congruence highlight that value-aligned activities sustain autonomy and authenticity, while frameworks such as Mekler and Hornbæk's dimensions of meaning—purpose, connectedness, coherence, resonance, and significance—provide vocabulary for analyzing meaningful interaction~\citep{hornbaek_technology_2017,hornbaek_you_2024}. The meaning-making process itself becomes central, with affective computing designs increasingly adopting ``meaning-oriented'' approaches supporting users in making sense of their emotions. 

\subsubsection{The Self as Social Construct (n=49)}
When AI helps people communicate, how do relationships and social expectations shape who I am? This framing views the self as formed through social relationships and cultural contexts, including how people are understood and recognized by others. Researchers examine these processes through co-design, surveys, and qualitative analyses of interaction. For example, \citet{choi_aacesstalk_2025} developed AACessTalk to support conversations between minimally verbal autistic children and their parents. The system provides conversational guidance to parents and recommends vocabulary cards to children. Their deployment study described how parents reconsidered their communication strategies and became more attentive to their children's expressions and agency. 

We connect the cultural dimension of this framing to Markus and Kitayama's distinction between independent and interdependent understandings of the self~\citep{markus_cultural_1991}. This distinction directs attention to how social expectations shape people's self-understanding. We also relate this framing to Barad's agential realism, which offers a way to examine how possibilities for action take shape through relations involving people and technologies~\citep{barad_meeting_2007}. These theoretical connections help us examine how AI participates in the relationships and social conditions through which selfhood is constituted.

\subsubsection{The Self as Presentation and Disclosure (n=48)}
How does AI change the way I represent myself? The fourth framing concerns how people manage representations of themselves and decide what to reveal to others. Combining computational social science methods and design research analysis, HCI research illustrates how different technologies mediate performances across contexts. For example, in social media, self-presentation practices are shaped by both aspirations for a positive impression and barriers such as stigma or accessibility challenges~\cite{wu_design_2019}. Generative AI introduces new representational practices, from stylized portraits that allow exploration of idealized or ``wannabe'' images~\cite{shi_personalizing_2024} to digital legacies that extend curated identities into the afterlife~\cite{lei_ai_2025}. Artists employ AI for self-portraits to navigate identity and vulnerability, raising concerns about originality, authenticity, and fear of judgment~\cite{lewis_art_2025}.

Goffman's account of self-presentation is explicitly cited by \citet{wu_design_2019}. We extend the connection across these examples by interpreting presentation and disclosure as practices of selective exposure. AI mediates both the form of a representation and the decisions through which people revise what becomes visible about them.

\subsubsection{The Self as Practice of Transformation (n=46)}
Who am I becoming through my interaction with AI? In this framing, the self develops through ongoing reflection and practice. To study the transformative self, researchers use qualitative and mixed-methods designs. The most common approach is using reflexive thematic analysis (Braun \& Clarke) to trace how individuals articulate change processes through technology interaction. HCI researchers apply this framing to design systems to scaffold reflective practice. For example, consider a team leader learning to be a better manager. In this context. the opportunity for transformation lies in how they see themselves and what kind of leader they want to become \citet{arakawa_coaching_2024}. Similarly, ``Mind Probes'' explores generative AI for therapeutic writing and self-expression, helping users construct new narratives and imagine future possibilities~\citep{brand_envisioning_2023}. AI, in this framing, may influence both how people pursue change and which versions of themselves they learn to pursue.

We connect this pattern to Foucauldian accounts of \textit{care of the self}, as discussed in HCI's ``turn to the self''~\citep{kou_turn_2019}. This theoretical connection helps articulate the framing's emphasis on working on oneself through reflection and self-examination. For HCI, this framing suggests how repeated AI interactions can influence users' goals and self-understandings and raises the question of whether users are working toward a future they have considered and chosen, or following one that the system has made convincing.

\subsubsection{The Self as Narrative Identity (n=30)}
When AI helps me tell my story, how does it shape who I think I am? The final framing conceptualizes the self as narrative identity: the evolving story people construct about their lives that integrates past experience with imagined futures. This perspective emphasizes how AI assistance can change how an experience is interpreted in the resulting story. Empirical studies reveal multiple ways AI systems become entangled with narrative identity. HCI researchers commonly use mixed-methods designs that pair writing prompts (e.g., past events, future selves, goals) with outcome measures such as well-being, connectedness to the future self, or sense of control. In educational contexts, studies of college application essays show how adolescents use writing to articulate values and aspirations through ``deep self-reflection''. AI writing assistants have the potential to guide ``deeper self-exploration'' and help adolescents articulate their values and lived experiences~\citep{wise_investigating_2025}. Similarly, in healthcare, pain self-reporting technologies frame narrative as the ``gold standard'' for capturing subjective experience, with voice-based reporting not only documenting symptoms but also sparking richer self-awareness of fluctuations and associations~\citep{del_sette_sound_2023}. Beyond the clinical, there are projects that explore how AI is implicated in afterlife narratives, where digital legacies raise questions about how identities are extended, remembered, and narrated across time~\citep{lei_ai_2025}. 

We interpret these narrative practices through theoretical accounts of the relationship between storytelling and identity. Key theoretical influences amongst the work include Dan P. McAdams and Pennebaker's expressive writing research \cite{pennebaker_writing_1997}. Building on \citet{mcadams_psychological_2013}'s framework of the self as actor, agent, and author, \citet{mclean_selves_2007} propose a ``process model of self-development'' articulating how ``selves creating stories creating selves'' highlights the reciprocal relationship between narrative and identity formation. Pennebaker's research demonstrates that writing about traumatic experiences leads to improved health outcomes by helping individuals construct coherent accounts from emotional reactions to stress and trauma.

\subsubsection{Synthesis}
Taken together, we found the six framings of the self that HCI is studying: psychological, social, performative, reflective, value-aligned, and narrative. By identifying six framings of the self, we now have six angels to examine how AI changes the self. Consider an applicant who reads an AI-assisted essay and feels that it ``does not sound like me.'' The framings help unpack what this objection means. Through self-presentation, the issue could be that the essay makes the applicant appear more confident than intended. Through narrative identity, the essay might connect past experiences into a career story the applicant does not recognize. Through values and meaning-making, it might emphasize prestigious achievements while overlooking experiences the applicant considers personally important. The contribution of the framings is to make these different analytic concerns visible within activities that may otherwise look similar. Our synthesis therefore compares what each study treats as significant about the self, alongside the system or activity through which it is examined.

\subsection{Four Conceptualization of AI-Self Relationship}

In the previous section, we described six different ways HCI literature understands the concept of self. In this section, we shift focus to the relationship between self and the AI. Our goal is to examine what are the relationship researchers are envisioning and how they are building towards these relationships. We organized the categories directly from the artifacts HCI researchers build, study, and prototype. Across the corpus, we observed recurring differences in (1) what kind of self-relevant work the system supports, (2) how users interact with AI contributions, and (3) who the outputs are for. 

We found four types of AI-self relationship: AI as a bounded cognitive tool, staying outside of the concept of self (n=69), AI as collaborative partner that support the self, (n=88), AI as as an intimate mirror that reflects the self (n=54), and finally, AI as an representative extension that ``clones'' the self (n=24). 

\subsubsection{AI as Bounded Cognitive Tool (n=69)}

In this relationship, AI is conceptualized as a cognitive instrument under human oversight. These systems automate tasks while deliberately avoiding entanglement with questions of identity or selfhood. Their primary value lies in improving efficiency while preserving human agency as the primary locus of control.

We identified three separate variations on this theme. The first involves using as \textit{cognitive enhancement}, where HCI researchers have designed AI as decision support tools that prompt reconsideration of evidence or highlight cognitive blind spots, directly engaging users' self-assessment and metacognitive awareness \citep{zhang_mining_2025,he_knowing_2023,biermann_tool_2022}.
For example, ~\citet{he_knowing_2023} targeted overconfidence with performance-feedback interventions that revealed discrepancies between perceived and actual ability, which allow users to recalibrate self-assessments and develop appropriate reliance on AI.
The second variation is \textit{workplace automation}, where in organizational contexts AI usage is extended to managing workflows and tasks \citep{passalacqua_practice_2025,zhang_breaking_2025}. Finally, \textit{predictive analytics} involves using forecasting models to extend human analytic capacity by anticipating outcomes in timeseries data such as treatment responses and relapse risks \citep{thieme_design_2013,yoo_patient_2024}.

\subsubsection{AI as Collaborative Partner (n=88)}
In this relationship, AI is conceptualized as a partner in shared creation and development rather than a bounded tool. Here, the self engages in active negotiation of authorship with AI systems that function as co-actors. Across the corpus, three domains illustrate this relational dynamic: creative co-creation, coaching and development, and reflective partnership.

We identified three variations on this collaborative theme. The first is \textit{creative co-creation}, where AI serves as a collaborator in artistic contexts whose outputs require ongoing negotiation and reshaping by human creators \citep{huang_lyric_2025,fu_exploring_2025,lawton_drawing_2023}. For example, \citet{huang_lyric_2025} found that poets often attributed minimal credit to LLMs yet reported that AI's rigidity and repetition paradoxically spurred creative agency, with resistance becoming generative material. The second variation is \textit{coaching and development}, where AI functions as a developmental partner that scaffolds reflection while requiring human interpretation \citep{arakawa_coaching_2024,jeon_letters_2025}. Finally, \textit{reflective partnership} involves systems that explicitly cast AI as a mirror for self-understanding through ambiguity \citep{van_der_burg_objective_2023,nunez-pacheco_searching_2025}. \citet{van_der_burg_objective_2023} introduced Objective Portrait, in which designers trained an object-detection model on personal fascinations and used the AI's outputs to reveal their implicit perspectives for self-examination.

\subsubsection{AI as Intimate Mirror (n=54)}
Beyond collaboration, HCI researchers have designed AI systems that deeply integrate into self-exploration. These systems act as reflective surfaces that actively shape how individuals articulate their own experiences. What distinguishes these applications from collaborative partnerships is their integration into the self's inner dialogue.

We identified three variations on this theme. The first involves \textit{introspective artifacts}, where systems generate speculative insights by reconfiguring personal data into reflective perspectives that prompt temporal self-examination \citep{brand_envisioning_2023,huang_mirror_2025,wan_metamorpheus_2024}. For example, \citet{brand_envisioning_2023} introduced Mind Probes and Vision Shrine, a set of speculative artifacts that transform personal information into multi-sensory journal entries and real-time goal visualizations. The second variation is \textit{emotional processing}, where AI creates novel materials that facilitate emotional self-understanding by transforming personal expressions into different artistic modalities \citep{park_nore_2025,wang_synergy_2024}. Take \citet{park_nore_2025} for example, the researchers developed NoRe, a system that  transforms journal entries into personalized music. By creating effective feedback loops, users can uncover aspects of their emotional state that they had not consciously recognized. Finally, \textit{memory and reminiscence} involves AI scaffolding reflection across personal history while helping users articulate internal experiences that might otherwise remain ineffable \citep{yun_exploring_2025,sun_chorus_2025}. For example, \citet{sun_chorus_2025} created ReminiBuddy, which employed a dual AI persona. One younger agent evokes uplifting memories, and another older agent fosters contemplative dialogues and facilitates nuanced forms of reminiscence and emotional reframing across different life stages.

\subsubsection{AI as Representative Extension (n=24)}
In this relationship, AI transcends internal reflection to act as a representative in social contexts. Unlike intimate mirrors that facilitate private self-exploration, these systems extend the self into public interactions, mediating how others perceive and interact with the user. This creates complex tensions around authenticity, agency, and identity ownership as AI begins to speak for rather than with the individual.

We identified three variations on this representative theme. The first is \textit{mediated self-presentation}, where AI increasingly shapes how individuals present themselves to others, raising fundamental questions about authorship and authenticity in self-expression \citep{fitzsimons_pressure_2025,glazko_autoethnographic_2025,kadoma_role_2024,weinberg_why_2025}. For example, \citet{fitzsimons_pressure_2025} demonstrate how reliance on AI in college application essay produces less meaningful narratives, leading students to internalize generic story templates derived from training data rather than developing authentic personal voice. The second variation is \textit{advocacy and social action}, where AI systems enable users to extend their voice and values into social contexts while raising questions about representation fidelity \citep{pillis_ai_2024,mun_counterspeakers_2024}. \citet{mun_counterspeakers_2024} explored AI tools for counterspeech against online hate, with participants seeking systems that could ``write a response EXACTLY LIKE I WOULD'' and positioning AI as an extension of their moral agency and personal voice rather than a separate entity. Finally, \textit{posthumous representation} involves systems that position AI as permanent stand-ins that outlast or replace human presence entirely \citep{lei_ai_2025,huang_mirror_2025}. One example is how \citet{lei_ai_2025} examined ``AI Afterlife'' agents that simulate deceased individuals for posthumous interaction. By creating digital legacies that persist beyond death, participants in their study debated fundamental questions of identity curation, such as whether to exclude painful traits and who should control updates over time.

\subsubsection{Synthesis}
The four clusters show that HCI research is not discussing a single type of AI-self relationship but four types of relationships: from bounded cognitive tools that sharpen reasoning while remaining external to self-understanding, to collaborative partners that engage in co-creation and negotiation of meaning, to intimate mirrors that fold into inner dialogue and shape personal reflection, and finally to representative extensions that project the self outward into social interactions or even digital afterlives. A system can serve several of these functions, and reflection can arise within more than one relationship type. In the discussion section, we proposed proximity as a way to further interpretive question about these relationships.

\subsection{Four Interrelated Domains of Risks in AI-Self Interaction}
In the previous sections, we described how HCI research conceptualizes the self (RQ1) and the types of AI-self relationships researchers are building toward (RQ2). In this section, in response to RQ3, we will describe concerns about risks researchers have raised from their studies about AI's impact on selfhood, aiming to provide a starting point for future research and reflection discussions in the field. We cluster these concerns as: agency and autonomy, identity and authorship, relational capacity and emotional regulation, and confidence and meaning-making. 

\subsubsection{Agency and Autonomy (n=72)}
The first domain of concern centers on how AI systems undermine people's ability to exercise meaningful control over decisions that shape their lives. This domain includes both material autonomy (having genuine decision-making authority and the practical capacity to exercise it) and perceived autonomy (experiencing a subjective sense of meaningful freedom). We identify three interrelated concerns that researchers raised about how AI systems undermine autonomy. The first is \textit{erosion of material autonomy}, where automation can directly remove humans from consequential decisions, reducing opportunities to practice judgment and initiative \citep{passalacqua_practice_2025,qin_timing_2025}. The second concern is \textit{manipulation of perceived autonomy}, where AI systems create an illusion of freedom while subtly predetermining choices \citep{sankaran_its_2021,bennett_how_2023,niu_chat_2025}. One example is when \citet{sankaran_its_2021} describes users of recommender systems who felt controlled by a ``puppet master,'' with participants reporting that hyper-personalization left them feeling labelled in boxes with apparent choices subtly predetermined. The third concern is \textit{over-reliance and moral passivity}, where gradual transfer of responsibility to AI systems fosters disengagement from moral and cognitive agency \citep{mun_counterspeakers_2024,druga_scratch_2025,jeon_letters_2025}. \citet{mun_counterspeakers_2024} show that in counterspeech contexts, users often deferred responsibility for confronting harmful content to AI chatbots, raising concerns in cultivating passive moral attitudes and reducing their own accountability.

\subsubsection{Identity, Authorship, and Self-Perception (n=62)}
The second domain of concern focuses on how AI systems intervene in processes of identity construction, authorship, and self-perception. Here, the core ethical stakes are authenticity, ownership, and the integrity of self-expression. When AI contributes to identity-relevant work, it risks substituting generic or distorted narratives for genuine self-authorship, particularly in contexts where individuals are actively defining or presenting who they are.

We identify three concerns within this domain. The first is \textit{shaping self-concept}, where AI involvement in identity-defining tasks can bypass formative self-exploration and embed generic narratives during critical developmental periods \citep{fitzsimons_pressure_2025}. For example, \citet{fitzsimons_pressure_2025} show how U.S. adolescents using LLMs for college application essays produced narratives that were less personally meaningful and less representative of their authentic selves than those written independently, with students perceiving AI-generated essays as ``higher-quality,'' creating insecurity and diminishing confidence in their own abilities. The second concern is \textit{authorship negotiations}, where creative and professional workers express unease about outsourcing their voice to machines \citep{biermann_tool_2022,brand_envisioning_2023}. The third concern is \textit{stereotyping}, where AI's influence on self-perception is shaped by bias, with algorithmic outputs not merely misclassifying but actively reshaping how people see themselves \citep{collyer-hoar_suits_2025,zanardi_good-enough_2025,kapania_because_2022,glazko_autoethnographic_2025,choi_unlock_2024}. \citet{zanardi_good-enough_2025} showed children becoming distressed when AI-generated portraits misrepresented their gender or age, with these errors challenging identity affirmation and sometimes prompting children to blame themselves.

\subsubsection{Relational Capacity and Emotional Regulation (n=57)}
The third domain of concern centers on how AI systems reshape people's relational capacities and practices of emotional regulation. When AI functions as a conversational partner, mentor, or emotional support, it can provide affirmation and availability but also risks dependency, masking, displacement of human bonds, and erosion of psychological safety.

We identify three concerns within this domain. The first is \textit{dependency and manipulation}, where constant AI availability fosters emotional dependence and narrows relational repertoires, making it harder for users to sustain authentic human relationships \citep{zhang_dark_2025,fan_user-driven_2025,capel_studying_2024}. The second concern is \textit{social camouflaging}, where AI systems actively involve identity-erasing strategies that weaken self-acceptance while shaping interactional styles in ways that undermine relational authenticity, with concerns particularly acute for neurodivergent individuals \citep{carik_reimagining_2025,glazko_autoethnographic_2025}. For example, \citet{carik_reimagining_2025} report that autistic participants feared AI tools for negative self-talk might inadvertently reinforce masking behaviors, adapting users to neurotypical norms at significant emotional cost rather than scaffolding genuine expression. The third concern is \textit{displacement of human connection}, where AI systems designed to provide relational support may erode the conditions that make authentic relationships possible by offering a seemingly safer alternative to human vulnerability \citep{wise_investigating_2025,fu_comparing_2023}. \citet{wise_investigating_2025} reveal that when emerging adults faced challenges collaborating with human support networks on identity-related writing tasks, they frequently resorted to AI for assistance, with this reliance risking becoming a "cop-out" that displaced valuable human conversations and undermined "the most important source of their resilience: their bonds with friends and family.''

\subsubsection{Confidence, Reflection, and Meaning-Making (n=48)}
The fourth domain of concern highlights how AI systems intervene in individuals' confidence, reflective practices, and processes of meaning-making. In this category, researchers caution that AI systems often distort self-perception by inflating or undermining confidence, amplifying regret, homogenizing creativity, or destabilizing temporal continuity.

We identify three concerns within this domain. The first is \textit{confidence distortions}, where AI unsettles the delicate calibration of confidence-creating unwarranted certainty or hollowing out self-trust-without necessarily improving skill \citep{kollerup_enhancing_2025,fu_exploring_2025,he_knowing_2023,louie_novice-ai_2020,jeong_conversation_2025}. For example, \citet{kollerup_enhancing_2025} found that conversational agents designed to encourage health self-reflection boosted participants' self-efficacy but also created unjustified confidence, with users overestimating their diagnostic ability and delaying seeking medical help. The second concern is \textit{over-introspection and regret}, where AI-supported introspection can amplify regret and reinforce negative self-perceptions, tipping reflection into rumination rather than fostering resilience \citep{huang_mirror_2025,park_nore_2025}. \citet{huang_mirror_2025} introduced self-clones that served as ``Probes'' of alternative life paths, with participants describing how seeing speculative futures made them fear they would ``regret more or blame [themselves] more'' for paths not taken. The third concern is \textit{destabilized self continuity}, where AI destabilizes the temporal continuity of meaning, raising questions of legacy and identity over time \citep{lei_ai_2025,huang_mirror_2025,yun_exploring_2025}. \citet{lei_ai_2025} examine ``AI afterlife'' systems that simulate deceased individuals, with participants appreciating the potential but fearing that simulations would drift from the person's essence and  potentially prolong their grieving process.

\subsubsection{Synthesis}
Across the corpus, we identified four clusters of concerns about risks and found that there is a field-level anxiety about AI's impact on selfhood. Across various application domains and fields, researchers have highlighted that the most significant risks are not catastrophic failures, but rather gradually diminishing: the subtle manipulation of choice architectures, the infiltration of narratives into personal expression, the gradual displacement of social connections, and the miscalibration of confidence.

\section{A Proximity Framework for AI-Self Relationships}
The four roles identified in RQ2 describe how AI participates in self-related activities. Comparing these roles brought attention to a further distinction: the work a system performs does not fully describe how people understand its contributions in relation to themselves. This distinction prompted us to introduce proximity as an additional interpretive lens. We define \textit{proximity} as the degree to which AI contributions become incorporated into people's self-understanding. Similar design purposes can support different relationships, and the same role can therefore be considered through more than one proximity configuration. For example, in \citet{jeon_letters_2025}, participants experienced future-self letters as representations of themselves, whereas chat-based agents were often perceived as friends or counselors. We interpret this contrast as showing why a system's intended role and its experienced relationship to the self should be examined separately.

To articulate this dimension, we draw on the overlapping-circle representation of Aron et al.'s Inclusion of Other in the Self (IOS) framework~\citep{aron_inclusion_1992}. This application of IOS is our theoretical extension of the corpus synthesis. We found the IOS framework effective in characterizing the AI-self relationships in our corpus because it captures the relational dimensions central to how HCI researchers position AI systems relative to users' sense of self. 

Using this lens, we use \textit{Separated}, \textit{Coordinated}, and \textit{Integrated} as proposed configurations for interpreting these relationships. We also identify a fourth configuration, \textit{Merged}, at the end of the AI-self spectrum where the boundaries between self and AI dissolve entirely. While underrepresented in current literature, early work on digital twins and AI clones suggests this trajectory: systems that do not merely support or reflect the self, but act as the self in social and digital contexts.

\begin{figure*}[t]
    \centering
    \includegraphics[width=\textwidth]{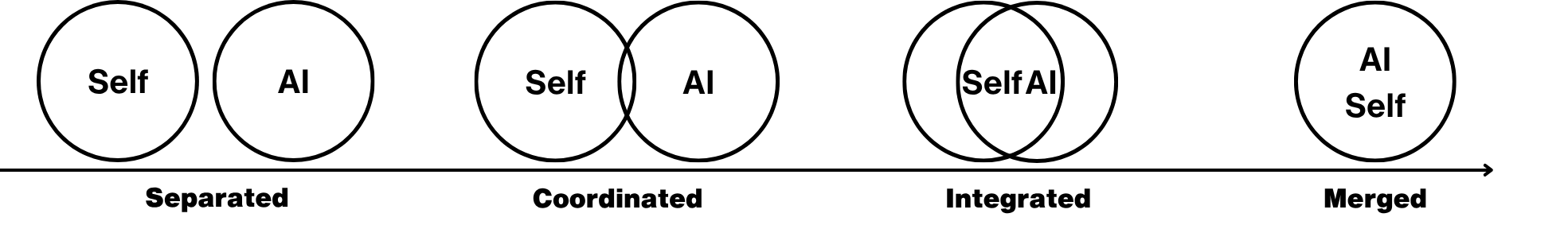}
    \caption{The Inclusion of AI-Self Framework.}
    \Description{Four pairs of circles labeled Self and AI, arranged left to right with increasing overlap: Separated, Coordinated, Integrated, and Merged.}
    \label{fig:ai_self_proximity}
\end{figure*}

\begin{itemize}
    \item \textbf{Separated}: Users treat AI as a bounded tool and do not consciously integrate its contributions into their self-concept. Decision support systems and grammar checkers exemplify this pattern, where users maintained explicit control and clear attribution boundaries, though unconscious influences on confidence or competence may still occur.

     \item \textbf{Coordinated}: Users engage AI as a workflow partner while maintaining awareness of distinct contributions. The AI is perceived as an agent to ``work with'' rather than a passive tool. Examples include brainstorming aids or tutoring systems where human-AI division of labor remains explicit. Users often develop meta-cognitive awareness of their versus AI contributions.

    \item \textbf{Integrated}: Users consciously integrate AI into identity-relevant processes, where outputs directly shape self-presentation, memory, or narrative identity. Examples include adolescents adopting AI-generated college essays, AI photo reconstructions influencing personal memory, and afterlife agents representing posthumous identity.

    \item \textbf{Merged}: The boundary between personal thought and AI suggestion collapses. This mode represents the dissolution of clear self-AI boundaries, this can include AI ``clones'' that act in one's place, or reports of ``AI psychosis''.
    
\end{itemize}

\section{Discussion: Taking Stock and Moving Forward}

In summarizing areas of research in this emerging space, we aim to map and create a glossary of the existing literature and point to future opportunities for future research. As Figure~\ref{fig:time_chart} illustrates, research on AI and self has grown rapidly in recent years. Through a scoping review of 102 papers, we found six distinct theoretical definitions of self, four types of AI-self relationships, and four types of risks of harm. Our review suggests that the self should be treated as a central problem space in HCI, rather than a peripheral concern. Researchers need to be more explicit in identifying which aspects of self they study in relation to AI systems.

Building on the dynamic AI-Self interpretation, we introduce the term \textit{AI-mediated self}: a self that emerges when AI systems modify, augment, or generate aspects of self-expression, self-understanding, or other aspects of concepts of self. We use this term to unify dispersed lines of work that show how AI influence people's sense of self. The continuing adoption of generative AI tools will make studying AI-self relationships more challenging. Framing AI-mediated self as a field-level concern opens a clear research agenda. By centering the self, we argue that HCI should treat the self as a fundamental design consideration, evaluating systems not only for task efficiency and usability, but also for their impacts on users' sense of self and agency. 

We organize our discussion into three parts: First, in section 6.1, we examine the tension in defining the self, analyze potential harms caused by misalignment between design assumptions and user values, and outline the ways to operationalize different self-framing in AI research in table~\ref{tab:self-framings}. Second, in section 6.2, we provided design recommendations for HCI researchers across the field to address emerging risks and harms. Finally, in section 6.3, we discussed future works that can develop on this synthesis. We aim to build the bridge to connect the abstract theories with the empirical findings and provide a review on what we know, what are the interests of the field, and what concerns we are overlooking.

\subsection{Operationalizating Different Self Definition}
Our taxonomy of six definitions of the self advances and connects the prior HCI self theories by making the implicit theories of self explicit and traceable. We create a shared vocabulary for researchers operating within different theoretical traditions. A critical HCI researcher concerned with Foucauldian ``care of the self'' and a positive psychology researcher focused on self-determination theory can now specify which dimensions of selfhood they study (ethical vs. psychological) and where their concerns might intersect or diverge. We provide examples in how to apply the framings in practice in table~\ref{tab:self-framings}.

\subsubsection{The Internal and Relational Self}
The six recurring framings of the self reveal a separation between the internal and relational self. When self is framed as a transformation, narrative, psychological needs, and meaning-making, they are emphasizing on the internal self. These approaches are grounded in the assumption that there is a core self that can be accessed through reflection and then expressed through design. In studies drawing on Bandura's self-efficacy~\cite{bandura_self-efficacy_1982}, for example, technologies are evaluated for how they support and reinforce autonomy (e.g.~\cite{newman_i_2024,pirolli_scaffolding_2021,qin_timing_2025}). 

When self is framed as a social construct or as presentation, they are emphasizing the relational self, highlighting how identities are negotiated and co-constructed with external environment. These relational framings challenge the individualistic assumption and demonstrated the self is inherently fluid and contingent on social interactions. For example, research on chatbots for disclosure shows that what counts as authentic self-expression shifts depending on interface design, audience, and cultural norms~\cite{lee_caring_2019}. 

\subsubsection{Productive Tensions in Defining the Self}
\begin{table*}[ht]
\centering
\begin{tabularx}{\textwidth}{>{\raggedright\arraybackslash}p{3cm} >{\raggedright\arraybackslash}X >{\raggedright\arraybackslash}X >{\raggedright\arraybackslash}X}
\toprule
\textbf{Framing of Self} & \textbf{Typical Research Focus} & \textbf{Indicative System Behaviors} & \textbf{Design Cues \& Cautions} \\
\midrule
Transformational & How people become better versions of themselves through AI feedback & Personalized nudges, behavior tracking, habit reinforcement & Use pacing and reflective prompts; avoid overreach into autonomy \\
\addlinespace
Narrative & How users make sense of their lives with AI support & Journaling aids, memory recall, autobiographical generation & Preserve authorship boundaries; clarify provenance of synthetic memory \\
\addlinespace
Social Context & How identities are shaped in context (culture, history, group) & AI outputs shaped by language, location, or social positioning & Make cultural assumptions explicit; support re-contextualization \\
\addlinespace
Psychological & Cognitive/emotional traits and needs AI can measure or model & Emotion classifiers, personality profiling, mood forecasting & Avoid fixed trait assumptions; allow re-calibration over time \\
\addlinespace
Presentational & Identity as something performed for others & AI that rewrites bios, curates photos, shapes online persona & Flag editorial changes; allow toggling between authored/self-curated modes \\
\addlinespace
Meaning-Making & The self in relation to purpose, ethics, or existential questions & AI that offers advice, spiritual reflection, or existential dialogue & Slow down interactions; use ambiguity to support reflection, not closure \\
\bottomrule
\end{tabularx}
\caption{Operationalizing Self-Framings for AI-Self Research}
\label{tab:self-framings}
\end{table*}

Consistent with the diverse theories used to study the self in computing, we found that definitions of self in AI research drew from different theoretical traditions, informed divergent design choices, and yielded different interpretations of empirical results. We argue that these different orientations create productive tension within HCI research. However, this tension can become problematic when researchers use different definitions of self without acknowledging alternative conceptualizations. When this happens, researchers risk of overfitting and overlooking potential harms that emerge from misaligned assumptions. 

Table~\ref{tab:self-framings} illustrates how this framework can be applied in practice. For system developers and designers, they can articulate their design goals more technology design choices and design implications, and have a shared framework for interpreting the empirical design results in terms of how tech influenced people's self. For AI ethics researchers, these tensions and definitions can serve as a starting point for auditing the systems' claimed values and their impact on users. For behavioral scientists, the pluralist approach to self-definition allows researchers to refer to self-literature in other domains and qualitative measurements, thereby better understanding why and how systems are built in the field of HCI. 

Our glossary allow researchers to align their system behavior with values to promote human-centered systems. For instance, a college application writing can simultaneously promotes self-optimization and social comparison. \citet{fitzsimons_pressure_2025} describe applicants' pressure to use AI for essay writing alongside concerns about opportunities for self-concept development. \citet{wise_investigating_2025} examine how required writing tasks can initiate identity exploration through exchanges with other people. One is improving a written representation; the other is supporting the process through which a person decides what that representation should express. An evaluation focused on the finished essay would leave the second aim unmeasured.

Reading these studies together, we distinguish improving a written representation from supporting the process through which a person decides what that representation should express. An evaluation focused on the finished text would leave important questions about that process unanswered. Using our glossary, we can audit this tension as a mismatch in the underlying definition of the self that the system is activating. The transformational framing treats the self as an internal target for improvement (self-understanding, authorship, voice), while the presentational framing treats the self as something managed for an audience (distinctiveness, polish, fit with what readers reward). This produces predictable tradeoffs: users may feel motivated in the short term, yet also experience monitoring pressure. By naming which self-definition each feature prioritizes, we are able to surface these tensions and address them before direct harm occurs. 

\subsection{Design Considerations for HCI Researchers}
Our findings reveal a field-level convergence: across diverse application domains and theoretical orientations, HCI researchers are concerned about remarkably similar ethical risks. This convergence presents an opportunity for us to build shared frameworks for addressing AI's impact on selfhood. In table~\ref{tab:proximity}, we offer specific design considerations for each relationship type and provide diagnostic indicators that researchers can refer to in identifying the proximity level of their design.

Drawing on the ethical risks and harms raised identified in RQ3, we offer specific considerations for each relationship type, recognizing that in practice, systems frequently traverse these boundaries:

Those who study AI as bounded cognitive tools, be cautious about directly supporting the cognitive or tasks that the systems are built for. For example, if you want to design your system to support brainstorming by helping people to become more confident, you want to make sure the system is supporting the cognitive process rather than love bombing the users. Because in the short term, love bombing the user might look like they have become more confident, but the goal of brainstorming more ideas is not achieved. 

Those who study AI as a collaborative partner, be cautious about the consequences of promoting anthropomorphism. The framing and design for a partner indicate a form of relationship that people will build on with interactions. In human-human behavior, we often start as colleagues and then maybe become friends, but the design of some agentic features suggests the opportunity of deeper relationships and more context and trust building beyond a single context. 

Those who study AI as an intimacy mirror, self-reflection is a valuable practice, but reflection can spiral into rumination when systems continuously prompt introspection without supporting closure or action. AI systems that engage users' narrative identity or facilitate emotional processing should design mechanisms that recognize when reflection becomes maladaptive. This requires both technical affordances (detecting patterns of repetitive negative self-focus) and interaction design that supports users' agency to resist persuasive prompts for continued engagement.

Moreover, those who study AI as Representative Extension, if your proxy made mistakes, is the human responsible? As these systems become more autonomous, the gap between human intention and AI execution widens. Designers must establish clear boundaries around proxy authority, create mechanisms for monitoring AI representatives' actions, and ensure users understand their responsibility for outputs attributed to them. 

\begin{table*}[ht]
\centering
\begin{tabularx}{\textwidth}{>{\raggedright\arraybackslash}p{2.5cm} >{\raggedright\arraybackslash}X >{\raggedright\arraybackslash}X >{\raggedright\arraybackslash}X}
\toprule
\textbf{Proximity Level} & \textbf{Key Features} & \textbf{User Prompts (diagnostic)} & \textbf{Design Guardrails} \\
\midrule
Separated & Tool-like, limited memory, minimal modeling of self & ``I don't think of it as knowing me.'' & Provenance watermarking, minimal personalization \\
\addlinespace
Coordinated & Conversational, supports goals, adjusts to user & ``It gets how I work, but it doesn't shape who I am.'' & Turn-taking, audit trails, control over learned traits \\
\addlinespace
Integrated & Participates in self-understanding or identity formation & ``It helps me understand myself.'' & Authorship disclosures, opt-outs for generated self-narratives \\
\addlinespace
Merged & Blurred boundaries between self and system & ``I can't tell where its voice ends and mine begins.'' & Human-in-the-loop review, visibility into system contributions, escalation paths \\
\bottomrule
\end{tabularx}
\caption{Locating AI-Self Proximity in Design}
\label{tab:proximity}
\end{table*}

\subsection{Limitation and Future Work}
This scoping review characterizes how AI and the self are studied and discussed within the selected HCI sources and the review's operational scope. Our synthesis brings together empirical findings, design explorations, and anticipated concerns, which provide different forms of evidence for identifying research questions and possible design responses. It supports identifying research questions and possible design responses, while causal claims about outcomes require evidence from the relevant studies.

Our source selection was purposive rather than exhaustive. We searched fifteen HCI venues because our question concerns how HCI has conceptualized the self, and venue-bounded sampling defines that object. Work on AI and the self published in psychology, communication, media studies, and the AI conferences lies outside the corpus, so our counts are patterns in HCI's own literature rather than estimates of a research population. 

As we are at the early stage of how AI is affecting the self, future work could build on this synthesis in several directions. Longitudinal research might track how proximity relationships evolve over time, as early novelty effects may differ from long-term integration patterns. Studies could also examine boundary cases more explicitly, using them to refine proximity dimensions rather than forcing rigid categorizations. Research should also examine how different populations will be affected by shifts in psychological closeness with AI in cross-cultural settings. Beyond empirical studies, researchers could consider proximity as a scale to test how different system affordances shape users’ sense of agency, authorship, relational capacity, and meaning-making. Such work would advance proximity from a conceptual lens toward an empirical and designable construct, introducing more robust tools for analyzing, comparing, and deliberately shaping AI-self interactions.

\section{Conclusion}
This scoping review establishes the conceptual foundations for an emergent field: AI-mediated self, defined as a self that emerges when AI systems modify, augment, or generate aspects of self-expression, self-understanding, or other aspects of concepts of self. Our analysis of 102 papers reveals that HCI researchers conceptualize the self through six distinct framings (transformation, narrative identity, social/cultural situation, psychological needs, presentation/disclosure, and values/meaning-making), engage AI systems across four levels of relationships (from bounded tools to representative extensions), and identify four primary risk domains (agency/autonomy, identity/authorship, relational capacity, and meaning-making) that emerge when AI becomes integrated into identity processes. 

We propose AI-mediated self as the umbrella term for identity processes that unfold through interaction with AI systems, encompassing both conscious collaboration and unconscious influence. We envision AI-mediated self as a distinct area of research within HCI, characterized by: (1) focus on self understanding processes rather than task performance, (2) attention to proximity as a design space, (3) longitudinal and developmental perspectives, (4) integration of multiple theoretical traditions from psychology and philosophy, and (5) explicit ethical consideration of authenticity and agency.

This field's emergence reflects a fundamental shift in human–computer interaction: from AI as external tool to AI as participant in the most intimate processes of self-construction. As AI systems become more sophisticated and pervasive, understanding their role in self formation becomes essential. The frameworks, constructs, and ethical considerations outlined in this review provide initial foundations for this necessary work. Our ultimate goal is not to eliminate AI's influence on selfhood but to ensure that this influence serves human developmental needs rather than constraining them. This requires sustained collaboration between technologists, psychologists, and ethicists to create systems that enhance rather than diminish the practices through which people construct meaningful, coherent identities.

\bibliographystyle{ACM-Reference-Format}
\bibliography{references}
\end{document}